\documentclass{elsart}
\usepackage{graphicx}
\usepackage{amssymb}
\begin{document}

%%%%%%%%%%%%%%%%%%%%%%%%%%%%%%%%%%%%%%%%

\begin{frontmatter}
\title{Effect of Neutral Current Interactions on High Energy
       Muon and Electron Neutrino Propagation through the Earth}
\author[label1]{Annalisa L'Abbate},
\ead{labbate@ba.infn.it}
\author[label1,label2]{Teresa Montaruli},
\ead{montaruli@ba.infn.it}
\author[label2,label3]{Igor Sokalski\corauthref{cor1}}
\ead{sokalski@ba.infn.it}
\corauth[cor1]{Corresponding author. Phone: +39-080-544-2347, 
               fax: +39-080-544-2470.}
\address[label1]{Physics Department, University of Bari, Via Amendola 173, 
                 I-70126 Bari, Italy}
\address[label2]{Istituto Nazionale di Fisica Nucleare / Sezione di Bari, Via 
                 Amendola 173, I-70126 Bari, Italy}
\address[label3]{Institute for Nuclear Research of the Russian Academy of 
                 Sciences, 60th October Anniversary Prospect 7a, RU-117312  
                 Moscow, Russia}

\begin{abstract}
High energy $\nu_{\mu}$ and $\nu_{e}$ propagation through the Earth has been 
performed using the Monte Carlo technique. We focused our attention on the 
effect of neutral current deep inelastic interactions 
$\nu_{l}\,N\stackrel{NC}{\longrightarrow}\nu_{l}\,X$
compared to that of charged current ones
$\nu_{l}\,N\stackrel{CC}{\longrightarrow}l\,X$. We have found that NCs do
not produce any significant effect with respect to the case in which only CCs 
are considered. Therefore we conclude that NC interactions can be neglected 
without considerable loss of accuracy. When computing upward-going neutrino 
fluxes a simple formula describing the transmission probability, 
that depends on 
the neutrino direction and energy and the CC cross section, can be used to 
account for the Earth shadowing effect.
\end{abstract}

\end{frontmatter}

%%%%%%%%%%%%%%%%%%%%%%%%%%

\section{Introduction}
\label{sec:intro}

Numerous works were published during the last decade on high energy neutrino
propagation through the Earth (see, {\it e.g.}
[1-11],
%\cite{gandhi1,naumov,stasto,dutta,beacom,new1,anis,bmss,new2,new3,new4}, 
though the whole list of
publications dedicated to this item is much longer). 
This is an important topic since, in order to correctly estimate the
event rates due to neutrinos crossing the Earth that interact close 
or inside underwater neutrino telescopes (UNTs)
%\cite{amanda,antares,baikal,icecube,nemo,nestor}, 
[12-17],
the Earth shadowing effect needs to be accounted for.

The most complex case is that of $\nu_{\tau}$. Besides suffering neutral 
current (NC) interactions, it undergoes a regeneration chain of processes in 
the Earth due to charged current (CC) interactions and consequent fast 
$\tau$-lepton decay, producing a $\nu_{\tau}$ of lower energy than the initial
one: 
$\nu_{\tau} \stackrel{CC}{\longrightarrow} \tau \rightarrow \nu_{\tau} \dots$.
Moreover, secondary $\nu_{e}$'s and $\nu_{\mu}$'s are generated via 
$\tau^{+}\rightarrow\nu_{e} \bar\nu_{\tau}\,e^{+}$, 
$\tau^{-}\rightarrow\bar\nu_{e} \nu_{\tau}\,e^{-}$, 
$\tau^{+}\rightarrow\nu_{\mu} \bar\nu_{\tau}\,\mu^{+}$ and 
$\tau^{-}\rightarrow\bar\nu_{\mu} \nu_{\tau}\,\mu^{-}$, 
decays. Hence, a $\nu_{\tau}$   
with initial energy at Earth of $E^{0}_{\nu_{\tau}}$ leaves it with an energy 
$E^{1}_{\nu_{\tau}}\le E^{0}_{\nu_{\tau}}$, being accompanied by some amount
of secondary $\nu_{\mu,e}$'s \cite{beacom,bmss}.

In contrast to $\nu_{\tau}$'s, $\nu_{e}$'s and $\nu_{\mu}$'s can only loose 
energy undergoing a NC interaction 
$\nu_{l}\,N\stackrel{NC}{\longrightarrow}\nu_{l}\,X$ or be absorbed due 
to a CC interaction 
$\nu_{l}\,N\stackrel{CC}{\longrightarrow} l \, X$~\footnote{We do not 
consider $W^{-}$ production through the resonant process 
$\bar \nu_{e}\,e^{-}\rightarrow W^{-} \rightarrow X$ at 
$E_{\bar \nu_{e}}\!\approx$\,6.3\,PeV (the Glashow resonance \cite{glashow1}) 
which occurs only to $\bar \nu_{e}$  in a very narrow energy range.}.
Generally, to account  properly these two processes one has to solve complex 
transport equations or apply Monte Carlo (MC) 
simulation. However, if NC interactions are neglected, the 
neutrino flux emerging from the Earth $\Phi_{1}(E_{\nu_{\mu,e}},\theta)$ is 
given by the following simple formula which depends only on the neutrino 
energy $E_{\nu_{e,\mu}}$ and nadir angle $\theta$:
\begin{equation}
\label{flux}
\Phi_{1}(E_{\nu_{\mu,e}},\theta)=S(E_{\nu_{\mu,e}},\theta)\times\Phi_{0}(E_{\nu_{\mu,e}},\theta). 
\end{equation}
Here $\Phi_{0}(E_{\nu_{\mu,e}},\theta)$ is the incident flux and 
$S(E_{\nu_{\mu,e}},\theta)$ is the screening factor (or transmission
probability) which is defined as \cite{gandhi1}:
\begin{equation}
\label{screen}
S(E_{\nu_{\mu,e}},\theta)=1-p_{abs}(E_{\nu_{\mu,e}},\theta)=\exp \left[-N_{A} \! \times \! \sigma_{CC}(E_{\nu_{\mu,e}}) \! \times \!\! \int \!  \rho (\theta, l) \, dl\right],
\end{equation}
where $p_{abs}(E_{\nu_{\mu,e}},\theta)$ is the absorption probability, 
$\sigma_{CC}(E_{\nu_{\mu,e}})$ is the total CC cross section, $N_{A}$ is the 
Avogadro number, $\rho(\theta, l)$ is the Earth density profile for a 
given direction $\theta$ and $l$ is the distance in the Earth on which
the integral is performed to calculate the column depth.

In this short paper we describe the results of the MC simulations for 
$\nu_{\mu}$ and $\nu_{e}$ propagation through the Earth, focusing on the role 
of NC interactions and analyzing the accuracy of the formula in 
Eq.~(\ref{flux}).

%%%%%%%%%%%%%%%%%%%%%%%%%%%%%

\section{Method and Results}
\label{sec:method}

\begin{figure}[htb]
\begin{center}
\includegraphics[height=7.0cm]{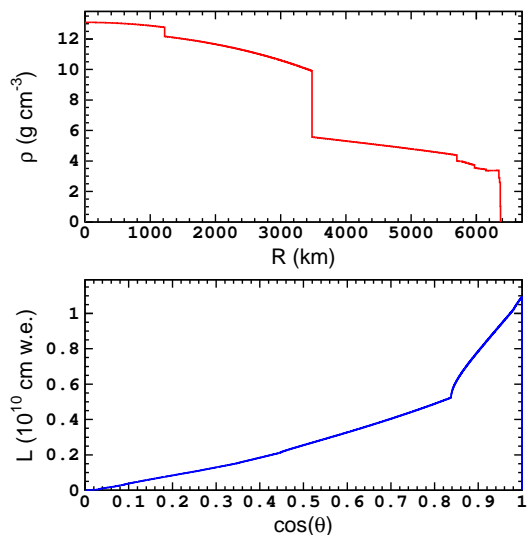}
\end{center}
\caption{\label{f0}
The Earth density profile \protect\cite{earth} (upper panel) and the Earth 
column depth vs nadir angle (lower panel) as it is used in our calculations. 
}
\end{figure}

The simulation we have used is fully described in \cite{bmss}. Shortly, we 
remind that deep inelastic cross sections are calculated using CTEQ3-DIS 
parton distribution functions \cite{CTEQ} taken from PDFLIB library 
\cite{PDFLIB}. We have assumed the Earth composition made by standard rock 
($A$=22, $Z$=11) of variable density with the Earth density profile taken 
from \cite{earth} (Fig.~\ref{f0}).

%%%%%%%%%%%%%%%%%%%%%%%%%%

\subsection{Results on monoenergetic neutrino beams}
\label{sec:mono}

We have simulated 1640 monoenergetic neutrino beams each of 5$\times$10$^{5}$ 
events ($\nu_{e}$, $\bar \nu_{e}$, $\nu_{\mu}$ and $\bar \nu_{\mu}$, 41 
neutrino energies in the range 10\,\,GeV $\le E_{\nu}\le$ 10$^{9}$\,GeV and
10 nadir angles in the range 0.1$\le\!\cos\theta\!\le$1.0) and we have saved 
information on neutrinos which survive after propagation.

\begin{figure}[p]
\begin{center}
\begin{tabular}{cc}
\includegraphics[width=6.6cm,height=7.1cm]{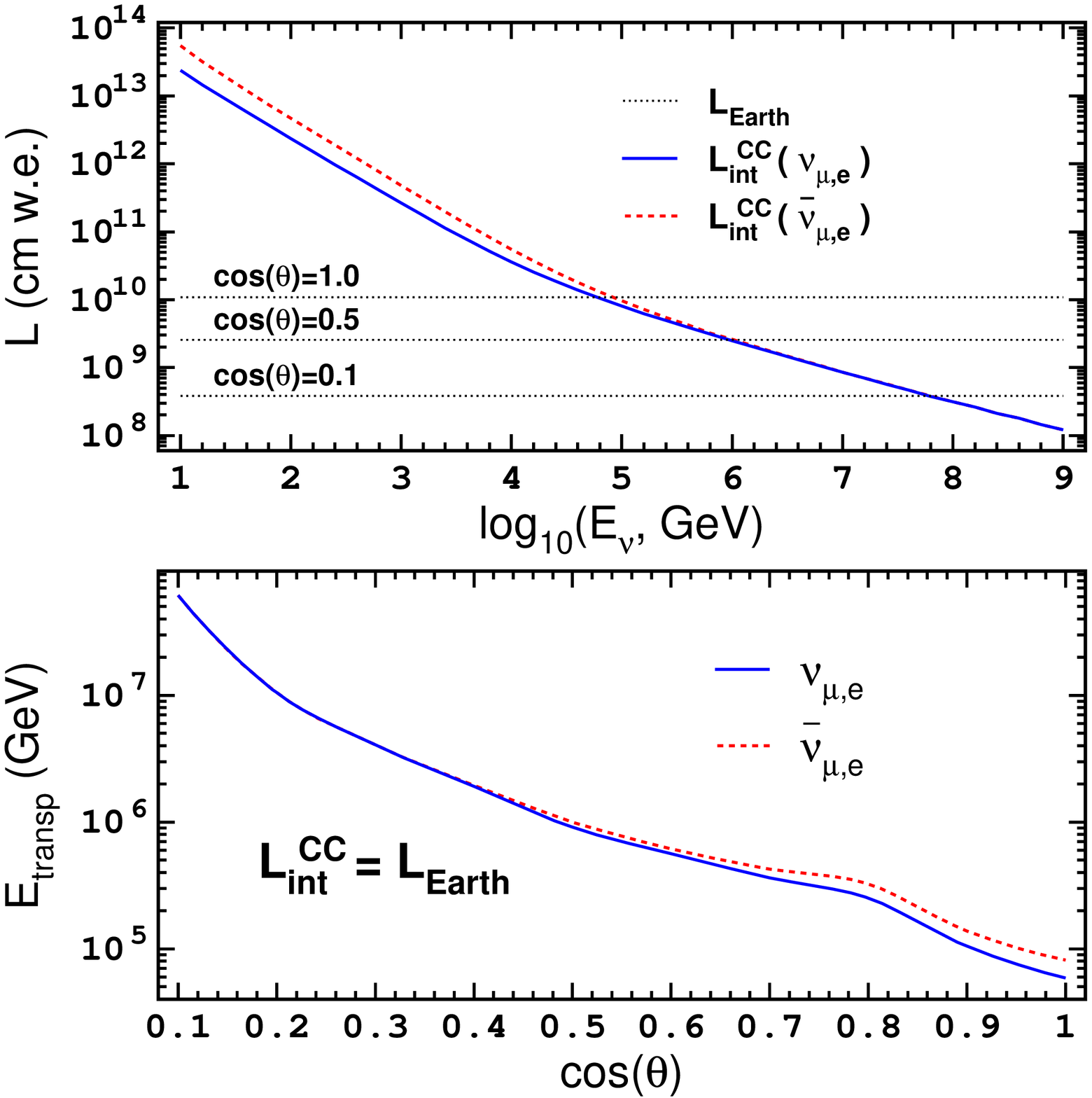}
&
\includegraphics[width=6.49cm,height=7.09cm]{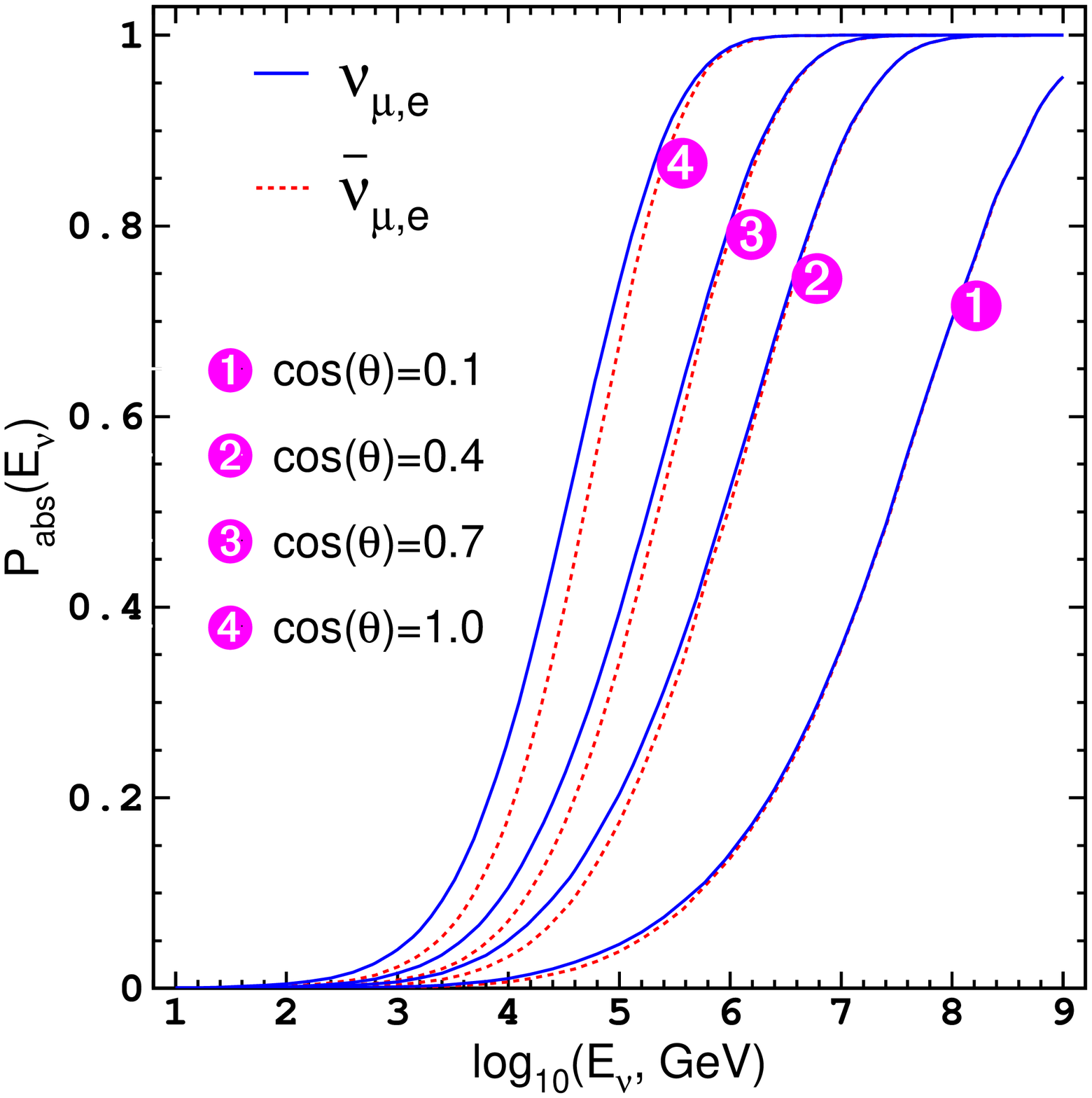}
\\
\end{tabular}
\end{center}
\caption{\label{f1}
Left upper panel: CC interaction length vs $E_{\nu_{\mu,e}}$. The three 
horizontal lines are the Earth column depth for 3 different values of the 
nadir angle ($\cos\theta = 0.1$, 0.5, 1.0). Left lower panel: transparency 
energy vs $\cos\theta$. The transparency energy is defined as the energy at 
which the neutrino interaction length equals the Earth column depth. 
A 'knee' at $\cos\theta \approx 0.8$ corresponds to an 'ankle' at the
curve for the Earth column depth vs nadir angle (Fig.~\protect\ref{f0})
which originates from the fact that the Earth density changes sharply
at the distance $\sim$3500\,km from the center.
Right panel: absorption probability for $\nu_{\mu,e}$'s ($\bar \nu_{\mu,e}$'s) 
in the Earth. In the simulation of propagation through the Earth only CC 
interactions were taken into account. In all panels solid lines are for 
neutrinos and dashed ones for anti-neutrinos.
}
\end{figure}
\begin{figure}[p]
\begin{center}
\begin{tabular}{cc}
\includegraphics[width=6.54cm,height=7.27cm]{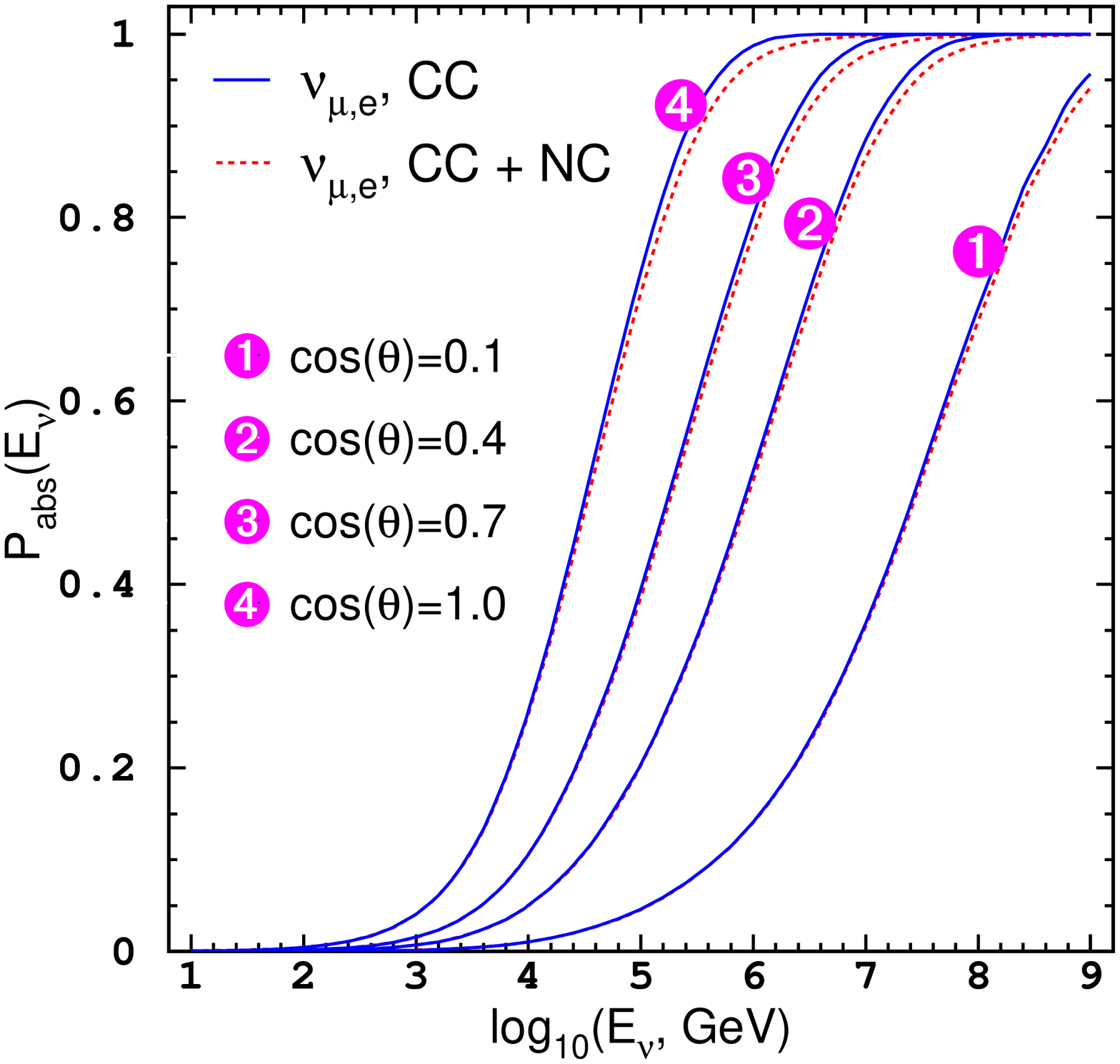}
&
\includegraphics[width=6.55cm,height=6.95cm]{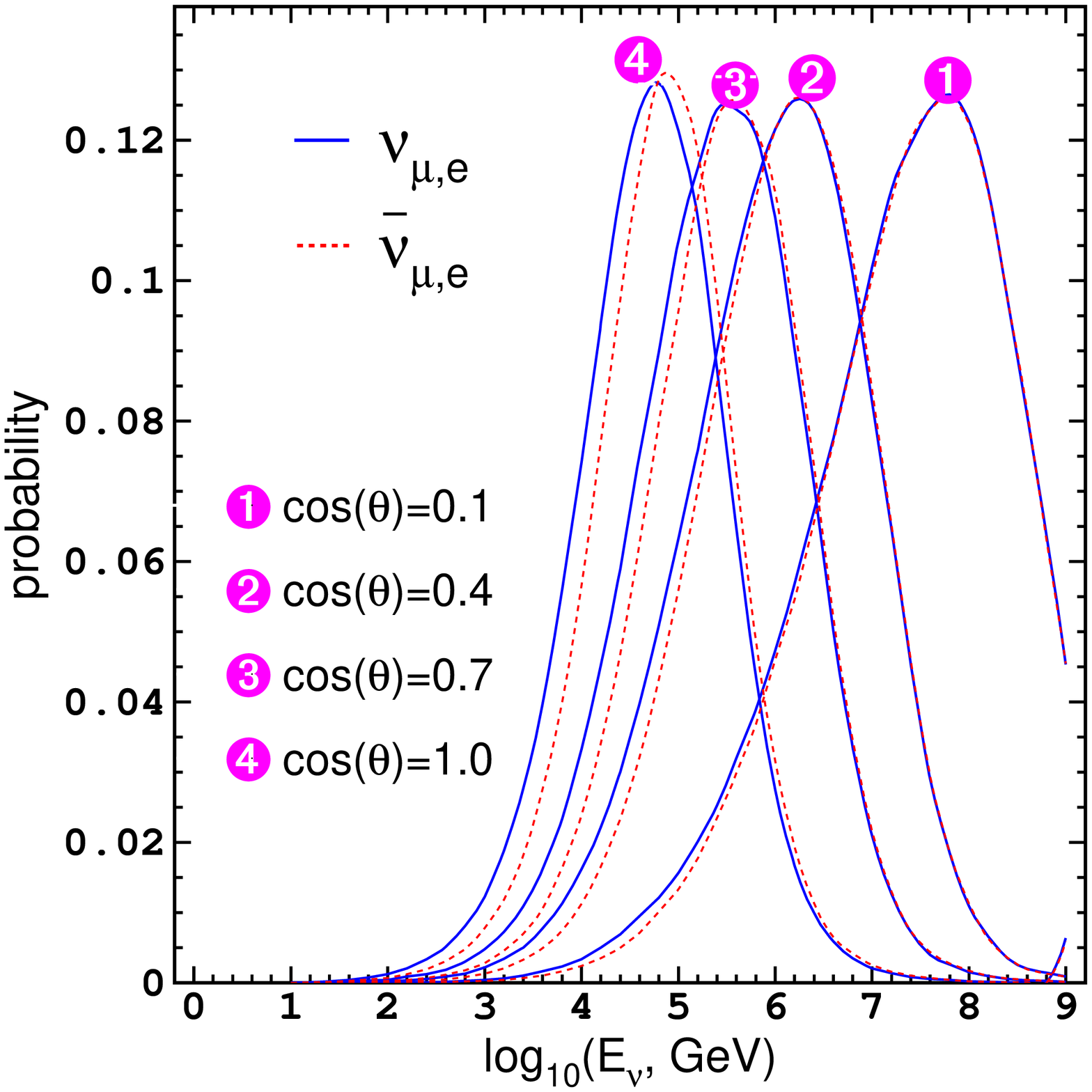}
\\
\end{tabular}
\end{center}
\caption{\label{f2}
Left panel: neutrino absorption probability in the Earth taking into account 
(dashed lines) or not (solid lines) NC interactions. Right panel: probability 
for $\nu_{\e,\mu}$'s ($\bar \nu_{\e,\mu}$'s) to pass through the Earth with no
CC interactions but with at least one NC interaction (solid lines are for 
neutrinos and dashed ones for anti-neutrinos). In both plots results are given
for four nadir angles ($\cos\theta = 0.1$, 0.4, 0.7, 1.0).
}
\end{figure}

The CC cross section $\sigma_{CC}$, that provides the probability of muon and 
electron neutrino absorption, grows with energy and consequently the 
interaction length  $L^{CC}_{int}=(N_{A}\times\sigma_{CC})^{-1}$ (in water 
equivalent units) decreases. At some 'transparency' energy $L^{CC}_{int}$ 
becomes equal to the Earth column depth seen by the neutrino for a given nadir
angle $\theta$ and so, at $E_{\nu_{\mu,e}}>E_{transp}(\theta)$ the Earth 
begins to consistently absorb $\nu_{\e,\mu}$'s. $E_{transp}(\theta)$ grows
with the nadir angle. It is about 50--80 TeV at directions close 
to the vertical 
($\theta=0^{\circ}$), it becomes about 60 PeV at $\cos(\theta)=0.1$ 
($\theta$=84$^{\circ}$) and increases to larger values for more
horizontal directions since the matter to be transversed decreases 
(Fig.~\ref{f1}, left panels).
For very horizontal directions the composition of the Cherenkov media 
where the array is located or also the details of the coast for
a specific underwater detector should be known to calculate
exactly the transparency energy. Anyway, its value will be very large.
Correspondingly, 
the probability that a neutrino undergoes a CC interaction and is absorbed 
grows with $E_{\nu_{\mu,e}}$. In the right panel of Fig.~\ref{f1} the 
absorption probability $p_{abs}(E_{\nu_{\mu,e}},\theta)$, computed as the 
ratio between the number of absorbed and propagated neutrinos for a given 
energy and nadir angle, is shown in the case where only CC interactions are 
simulated. 

In the considered energy range, $\sigma_{NC}$ is 2.3--3.2 times smaller 
compared to $\sigma_{CC}$ and hence, the probability that a CC interaction 
occurs is higher than that of a NC. When $\nu_{\e,\mu}$'s interact by a NC, 
the neutrino energy decreases on average by a factor 
$1-\langle y\rangle=$\,0.5--0.8 (depending on energy), since the rest of the 
energy 
$\langle y\rangle E_{\nu_{\mu,e}}$ is taken by the hadronic shower. Hence, the
probability that another CC interaction occurs decreases. In other words, the 
probability $p_{abs}(E_{\nu_{\mu,e}},\theta)$ that the neutrino will be 
absorbed becomes lower. But quantitatively the effect is very small as it can 
be seen in the left panel of Fig.~\ref{f2} where 
$p_{abs}(E_{\nu_{\mu,e}},\theta)$ for $\nu_{\e,\mu}$'s is plotted for 
propagation in the Earth with- and without accounting for NC interactions 
(results for  $\bar \nu_{\e,\mu}$'s are qualitatively the same). Comparing the
NC$+$CC case to the CC only case, we observe only a slight decrease of the 
absorption probability. The right panel of Fig.~\ref{f2} shows the probability
$p_{NC}$ that $\nu_{\e,\mu}$'s survive after crossing the Earth with no CC 
interactions but at least one NC interaction. The maximum value of $p_{NC}$ 
for different neutrino nadir angles is about 0.13 for energies around the 
transparency energy $E_{transp}(\theta)$. At lower energies it falls down due 
to the cross section $\sigma_{NC}$ almost linear decrease with energy. 
At higher energies it is 
suppressed due to the increase of $\sigma_{CC}$ since, even after loosing 
energy in NC interactions, the neutrino energy remains high enough to undergo 
another CC interaction that makes the neutrino disappear. In any case, 
$p_{NC}$ is remarkably lower compared to $p_{CC}$ and, therefore, it does not 
affect significantly the total $p_{abs}(E_{\nu_{\mu,e}},\theta)$, which turns 
out to be very close to $p_{CC}$.

%%%%%%%%%%%%%%%%%%%%%%%%%%%

\subsection{Results on various energy spectra}
\label{sec:spectra}

\begin{figure}[htb]
\begin{center}
\includegraphics[width=13.1cm,height=9.4cm]{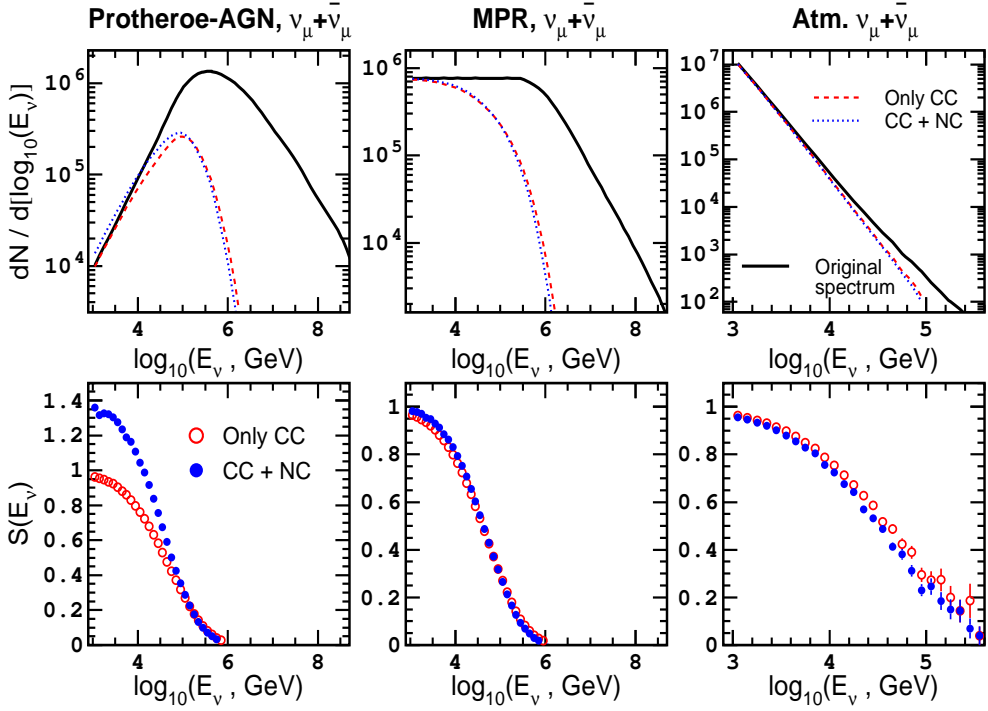}
\end{center}
\caption{\label{f3}
Upper row: muon neutrino spectra before (solid line) and after propagation 
through the Earth in the vertical direction with- (dotted lines) and without 
(dashed lines) accounting for NC interactions. Left: AGN Protheroe spectrum 
\cite{proth}; middle: MPR upper bound on diffuse neutrino flux \cite{mpr}; 
right: atmospheric muon neutrino spectrum in \cite{naumov1} including both 
neutrinos from pion and kaon decay and prompt neutrinos from charmed meson 
decay (calculated in the Recombination Quark-Parton Model frame). We 
considered the same amount of neutrinos and anti-neutrinos
($\phi_{\nu_{\mu}}\!:\!\phi_{\bar \nu_{\mu}}\!=\!1\!:\!1$). Lower row: the 
screening factor 
$S(E_{\nu_{\mu}})=\Phi_1(E_{\nu_{\mu}})/\Phi_0(E_{\nu_{\mu}})$ computed with 
fluxes presented in upper plots. Open circles: accounting for CC interactions 
only; full circles: MC including both NC and CC interactions. 
}
\end{figure}

We have investigated the effect of muon neutrino propagation through the 
Earth~\footnote{All the results and conclusions relevant to $\nu_{\mu}$'s are 
also valid for $\nu_{e}$'s since differential cross sections are practically 
the same.} for various models of neutrino fluxes incident on the Earth when 
only CCs are simulated and when also NCs interactions are accounted for. 

We considered the spectrum predicted by Protheroe \cite{proth} for 
diffuse fluxes of neutrinos produced by active galactic nuclei (AGNs), a flux 
equal to the Mannheim, Protheroe and Rachen (MPR) upper bound \cite{mpr} and 
the atmospheric muon neutrino spectrum from \cite{naumov1}. 
Results are shown in 
Fig.~\ref{f3} only for the vertical direction, that is the direction where 
interactions are most probable given the largest column depth. 
Results on screening factors (shown in the three lower panels) 
are in agreement 
with the results reported in \cite{stasto},
but, as we will see below, large values of screening factors ($S>1$) cannot 
be interpreted as a significant enhancement of the 
emerging neutrino spectrum due
to NC interactions (as it is concluded in 
\cite{stasto}). As a matter of fact, this effect is due to the spectral shape.
Considering upper panels in Fig.~\ref{f3} where emerging neutrino spectra are 
given one can conclude that effects of NC interactions can be summarized in
two points. Firstly, NCs result in a slight 
shift of the mean energy of the spectra emerging after 
propagation to lower energies compared to the spectra obtained simulating only
CCs (the spectral shapes are almost the same). This energy shift decreases 
with increasing nadir angles $\theta$. For the AGN spectrum \cite{proth} 
is equal to 22\%, 9\% and 3\% for $\theta$=0$^{\circ}$, $\theta$=60$^{\circ}$ 
and $\theta$=80$^{\circ}$, respectively. The shift is due to the decrease
of neutrino energy in NCs since a fraction of the parent neutrino
energy is taken away by the hadronic shower, hence the resulting neutrino
has lower energy than its parent.
Secondly, some enhancement of emerging spectra is observed when NC 
interactions are accounted for with respect to the case in which the 
simulation is performed without NCs. The most remarkable enhancement is found 
for the AGN spectrum in Ref.~\cite{proth} for 
which the ratios of the integral of the 
emerging spectra with and without including NC interactions are equal to 1.13,
1.02 and 1.003 for $\theta$=0$^{\circ}$, $\theta$=60$^{\circ}$ and 
$\theta$=80$^{\circ}$, respectively. 
Using the values for 
$\theta$=60$^{\circ}$ ($\cos\theta=$\,0.5) and this AGN flux to estimate 
the corresponding 
effect for the whole lower hemisphere (given that dependencies are more or 
less linear ones) we obtain a 9\% energy shift in the
spectral maximum and a 2\% flux enhancement 
due to NC interactions. Thus, when NCs are taken into account, neutrino 
spectra are shifted to lower energies by an amount that hardly will be 
appreciated by UNTs for which the typical error on neutrino energy 
measurement is never less than a factor of 2 (see, {\it e.g.}, \cite{ant1}). 
Also the enhancement 
of neutrino fluxes (and consequently of counting rates) due to NCs is small 
compared to experimental and simulation uncertainties in UNTs
[12-17].
%\cite{amanda,antares,baikal,icecube,nemo,nestor}. 

The noticeable increase (up to $S(E_{\nu_{\mu}})\approx$\,1.35) of the 
screening factor in the left bottom panel of Fig.~\ref{f3} for the 
considered AGN 
spectrum \cite{proth} in the TeV range due to NC interactions is in a good 
agreement with \cite{stasto}. But we believe that $S(E_{\nu_{\mu}})$ is not 
the proper variable to estimate the effective enhancement of the emerging 
spectrum after propagation, and hence of event rates. The effect of NCs
can be understood better if we compare the neutrino emerging
spectra obtained with and without accounting for NCs. As shown in 
Fig.~\ref{f3} (upper panels) these spectra are very similar. Hence, the 
enhancement in terms of $S(E_{\nu_{\mu}})$ is the result only of the slight 
energy 
shift of the emerging spectrum after propagation in the Earth relative to the 
initial spectrum at the Earth surface. 
The increase on the screening factor is larger for spectra with a maximum
and steeply falling down for decreasing energies below the energy 
corresponding to the maximum (see Protheroe AGN spectrum in Fig.~\ref{f3}). 
The effect is much smaller for flat spectra or 
power law spectra with negative spectral index, such is the case for 
MPR and for atmospheric neutrino spectra. 

%%%%%%%%%%%%%%%%%%%%%%%%%%%%

\section{Conclusions}
\label{sec:conclusions}

We have shown that the effect of NC interactions of ultra and extremely high 
energy muon and electron neutrinos propagating through the Earth on the  
neutrino flux emerging from the Earth does not play a significant role.
Consequently event rates in UNTs,
such as AMANDA \cite{amanda}, ANTARES \cite{antares}, Baikal \cite{baikal},
IceCube \cite{icecube}, NEMO \cite{nemo}, NESTOR \cite{nestor},
are not much affected by NC effects during propagation in the Earth.
Their effect can be neglected if the required accuracy
of calculations is lower than $\approx$10\% in terms of energy measurement and
few percents in terms of absolute event rates. Calculations of counting rates
can be much faster if the simple formula in Eq.~(\ref{flux}) is used compared 
to a full propagation of neutrinos through the Earth with MC simulations or 
semi-analytical methods.

%%%%%%%%%%%%%%%%%%%%%%%%%%%%

%%%%%%%%%%%%%%%%%%%%%%

\end{document}